# Experimental evidence for the separability of compound-nucleus and fragment properties in fission

Karl-Heinz Schmidt, Aleksandra Kelić, Maria Valentina Ricciardi

Gesellschaft für Schwerionenphysik, Planckstr. 1, 64291 Darmstadt, Germany

**Abstract:** The large body of experimental data on nuclear fission is analyzed with a semi-empirical ordering scheme based on the macro-microscopic approach and the separability of compound-nucleus and fragment properties on the fission path. We apply the statistical model to the non-equilibrium descent from saddle to scission, taking the influence of dynamics into account by an early freeze out. The present approach reveals a large portion of common features behind the variety of the complex observations made for the different systems. General implications for out-of-equilibrium processes are mentioned.

*Introduction*

Nuclear fission is a prominent example for the decay of a meta-stable state, in which many properties of the decay process are determined on the out-of-equilibrium descent outside the meta-stable state beyond the barrier. Out-of-equilibrium processes still pose a severe challenge to their theoretical description. Therefore, it is not surprising that the realistic modelling of the fission process with all its facets is far from being achieved at present times. This situation is particularly unsatisfactory since nuclear fission plays an important role in technical applications, like energy production or secondary-beam facilities, and in other fields of physics, like the nucleo-synthesis in the astrophysical r process [1].

The importance of nuclear fission for fundamental research on out-of-equilibrium physics is favoured by the wealth of observables it provides, the most relevant ones being the split of the fissioning system in mass and charge, the kinetic energies of the fragments and the amount and the energy distributions of neutrons and gammas emitted from the fragments.

*Brief review on experimental knowledge and theoretical models*

During almost seven decades of research, an immense body of experimental data on fission has been accumulated. In addition, tremendous effort has been invested on its theoretical understanding. Nevertheless, due to the experimental limitations on one side, and the difficulties in the theoretical description on the other side, the full understanding of the fission process has still not been reached. In the following we will first summarize the main features of the available experimental results and review the main lines of the different theoretical models and ideas developed in the past.

The identification of the barium isotopes $^{139,140}$Ba as decay products of the reaction $^{nat}$U + n, which represented the discovery of nuclear fission [2, 3], was already a hint for the predominant split into two fragments of different size. This feature remained a puzzle for some time, because it could not be explained by the liquid-drop model [3,4,5]. Kinematical experiments [6,7] confirmed the dominance of mass-asymmetric splits in low-energy fission of most of the actinides. After the formulation of the shell model [8,9], this peculiarity was attributed to shell effects in the fragments, namely the $N$=82 and $Z$=50 shells realized in the doubly magic $^{132}$Sn nucleus, which was assumed to form a stable cluster that fully survived in the heavy fragment [10]. With the postulation of deformed shells by Nilsson [11] and the macro-microscopic approach of Strutinski [12], also shell effects at large deformation were considered [13]. After first ideas to explain the mass distributions in fission by the influence of shells in the fragments by Fong [14] and Ignatyuk [15], Wilkins, Steinberg and Chasman [16] made a quantitative prediction on the mass distributions in fission on the basis of their statistical scission-point model, considering the influence of spherical and deformed neutron shells in the fragments on an equal footing. Their model had a remarkable success in qualitatively explaining the structural features in fission, although there remained important quantitative discrepancies. Theoretical calculations, e.g. with the two-centre shell model [17,18,19], revealed the microscopic structure of the whole potential-energy surface as a function of elongation and several other shape degrees of freedom like necking, triaxiality and mass asymmetry. But considering independent deformations of both nascent fragments in two-centre shell-model calculations is still a challenge [20]. Experimentally the concept of fission modes was introduced [21]. Different modes were attributed to different components in the two-dimensional mass-TKE distributions. Calculations of the shell structure on the fission path were quite successful in qualitatively explaining the features of multi-modal fission [22,23,24]. Still the integrity of the spherical shells ($N$=82, $Z$=50 and $N$=50) continued to be postulated [25,26]. It also plays a role in the random neck-rupture model of Brosa [22]. The importance of $^{78}$Ni and $^{132}$Sn clusters was also extracted from the steep decrease of the yields to lower masses in the light and the heavy fragments, respectively [27,28].

Only a few attempts were made to quantitatively predict the fission-fragment yields on a theoretical basis. In contrast to the scission-point model of Wilkins *et al*. [16], mentioned above, Dujvestijn *et al*. [29] calculated the mass yields in a statistical approach on the basis of the potential-energy landscape at the outer saddle. Both models are based on the macro-microscopic approach. First calculations with a microscopic approach, based on time-dependent Hartree-Fock calculations using the generator coordinator method [30], have been performed for the low-energy fission of $^{238}$U. The calculated fission-fragment kinetic energies and mass distribution [30] agreed fairly well with the experimental data.

The present experimental knowledge on shell structure in fission-fragment yields is illustrated in Figure 1. The figure shows mass distributions from particle-induced fission of stable or long-lived fissile targets, spontaneous fission of nuclei produced by heavy-ion fusion or breading, and element distributions from the electromagnetic-induced fission of secondary projectiles. The latter stem from one experiment performed in inverse kinematics at GSI, Darmstadt [31]. The distributions can roughly be classified as a function of the mass of the fissioning system: There is a gradual transition from a single Gaussian to a double-humped distribution around $A$=226, with triple-humped distributions appearing in the transition region. Above *A*=257, the distribution changes abruptly to a narrow symmetric one. Strong signatures of nuclear structure are also found in the TKE, the mass-dependent neutron-emission yields and other observables. These dominant global features can be associated to 4 fission channels:

1. symmetric fission of lighter nuclei,
2. asymmetric fission with a heavy almost spherical fragment around *A*=132 and a strongly deformed light fragment,
3. asymmetric fission with a strongly deformed heavy fragment around *A*=140, and
4. symmetric fission with two almost spherical fragments around *A*=132.

Indications for still another very asymmetric fission channel with a light fragment around $N \approx 52$ [32] and slight modulations of symmetric fission as a function of mass asymmetry were also reported [33].

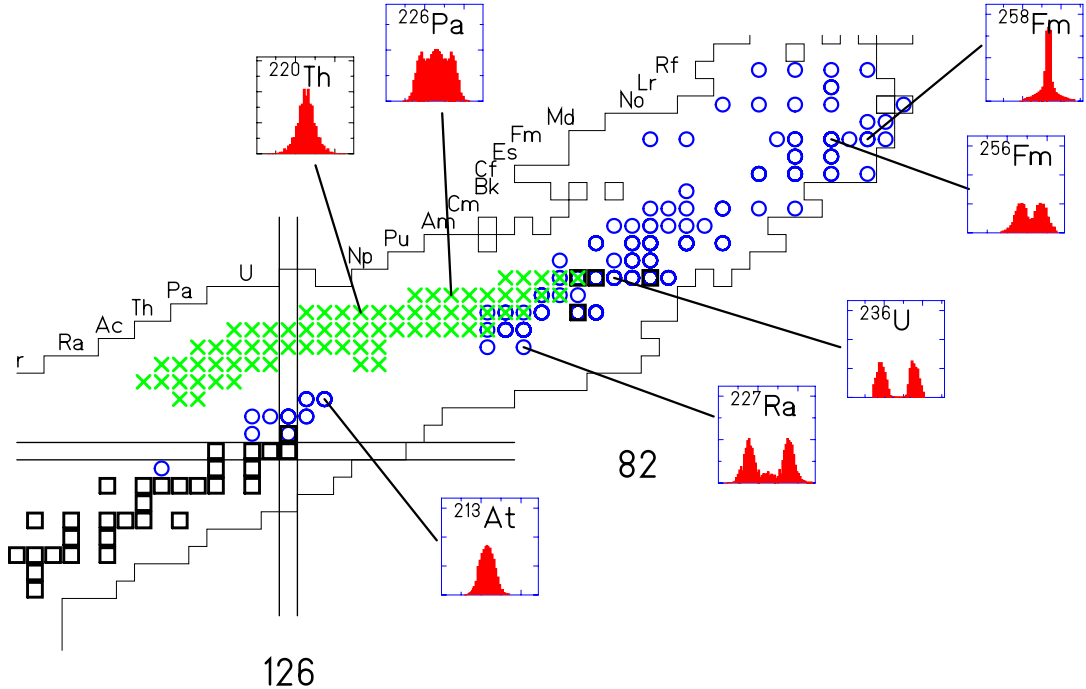

**Figure 1: (Colour online) Systematic overview on the structural features in low-energy fission of heavy nuclei. Circles indicate those nuclei for which mass distributions have been measured. Crosses mark the secondary projectiles for which element distributions after electromagnetic-induced fission have been obtained [31]. The insets show the mass, respectively element, distributions of the fission fragments. See [31] for references of the data.**

*The separability principle as an ordering scheme for fission*

Many extremely valuable and fruitful ideas have constantly contributed to improving our understanding of nuclear fission. Still, a realistic modelling of the fission process with all its facets is far from being achieved at present times. In this section, we will try to evaluate the potential of the available ideas and approaches by combining a number of those in the most efficient way. The separability principle of the compound-nucleus properties and the properties of the nascent fragments plays a key role in these considerations. It is established on the basis of the macro-microscopic approach, exploiting specific features of the microscopic potential.

Since the phase space acts as the most important driving force in almost any kind of physical process, we take the statistical model as the basis of our considerations. However, it is not trivial how to apply the statistical model to an out-of-equilibrium process. This should still be possible, because the fission process is characterized by an instability in only one dimension, i.e. the fission direction. All other degrees of freedom are stabilized by high walls around the fission path. However, the configuration at which the statistical model should be applied needs further consideration. Bohr and Wheeler introduced the transition-state concept to deduce the fission probability from the number of states at the fission barrier. Also the K quantum number, the projection of the angular momentum on the symmetry axis, is determined at saddle in light-particle-induced fission [34,35]. But other properties, like the mass distribution or the total kinetic energy might be determined at different moments. Indeed, it is the time scale of the process that is responsible for the evolution of the degree of freedom connected with a specific observable in comparison with the dynamical time of the fission process, which determines the configuration that is most relevant for this observable. It has been deduced from Langevin calculations that the time scale of the mass-asymmetry degree of freedom is comparable or slow compared to the saddle-to-scission time [36], and thus the decision on the mass distribution should be made quite early, not far beyond the outer saddle. The $N/Z$ degree of freedom, however, has been shown to evolve much faster [37], indicating that the decision on the charge density is made close to scission.

The important aspect of applying the statistical model to a specific configuration on the fission path is that it offers the possibility to determine the relevant properties of the potential-energy surface in this specific configuration from experimental data. This approach had been applied in refs. [33,38,39] to deduce the mass-asymmetric potential from measured mass distributions. At high energies, when shell effects have washed out, this procedure yields the stiffness of the macroscopic potential. By analysing the body of available data, the authors of refs. [33,39] arrived at a global overview, which they parameterised as a function of the fissility of the system.

Starting from this, it is convenient to apply the macro-microscopic approach for analysing the microscopic corrections to the macroscopic potential from the appearance of fission channels. This procedure was introduced in ref. [38], where the nuclear deformation potential energy and its components – the macroscopic and the shell correction part – have been extracted from experimental fission-fragment mass distributions for nuclei lighter than thorium. Thus obtained empirical shell corrections were in a satisfactory agreement with theoretical predictions [40]. Moreover, in the investigated region of fissioning nuclei, the extracted shell corrections showed almost no variation with the mass number of the fissioning nucleus, which lead the authors of ref.

[38] to conclude that shell corrections have "universal" character. We decided to extend this procedure and to apply it to a much larger range of fissioning nuclei, in order to test if the universality of shell corrections is still valid. Another difference to the work presented in the ref. [38] is that we consider the shell corrections in neutron and proton number, as expected from theory [16,17,18], and not in mass as done in [38]. We applied this procedure for $^{226}$Th, $^{239}$U, $^{252}$Cf and $^{260}$Md as shown in Figure 2. The results as a function of atomic number, respectively mass number, are also projected on neutron number. For this illustration of the method, neutron evaporation and charge polarization have been neglected.

The shell effects in the configuration, which is relevant for the mass split, were deduced by assuming a constant-temperature level density. For fission below the barrier, the mass distribution is not determined by the phase space but by the variation of the tunnelling probability through the outer barrier as a function of mass asymmetry. This is relevant for the spontaneous fission of $^{252}$Cf and $^{260}$Md. Also the fission of $^{239}$U, formed by capture of 1.7 MeV neutrons in $^{238}$U, is mostly governed by tunnelling, since the height of the conditional saddle is higher than the excitation energy of the compound nucleus for most mass splits. The tunnelling probability was calculated with the Hill-Wheeler approach. Parameter values used for the calculations are given in Table 1.

For $^{226}$Th, the central part of the charge distribution is explained in its shape by the macroscopic potential. Therefore, we assume the shell effect in this region to be zero. For $^{239}$U, the minimum at symmetry is determined by the macroscopic potential. For $^{252}$Cf and $^{260}$Md, where this kind of normalization is not applicable, we fixed the absolute value of the shell effect at $N$=90 to be the same for the other systems as for $^{239}$U. As expected, the microscopic structure turns out to be as complex as the charge or mass distributions.

Table 1: Standard deviation of the macroscopic mass distribution and effective temperature used for determining the shell-correction energies (Figure 2). For spontaneous fission, the oscillator energy of the inverted parabola of the outer barrier is given by $\hbar\omega = 2\pi \cdot T_{eff}$.

| System | $\sigma_A$(macroscopic) | $T_{eff}$ |
|---|---|---|
| $^{226}$Th (E* ≈ 11 MeV) | 8.8 | 0.6 MeV |
| $^{238}$U(n,f), $E_n$ = 1.7 MeV | 9.5 | 0.4 MeV |
| $^{252}$Cf (spont. fission) | 11.3 | 0.6 MeV |
| $^{260}$Md (spont. fission) | 12.2 | 0.6 MeV |

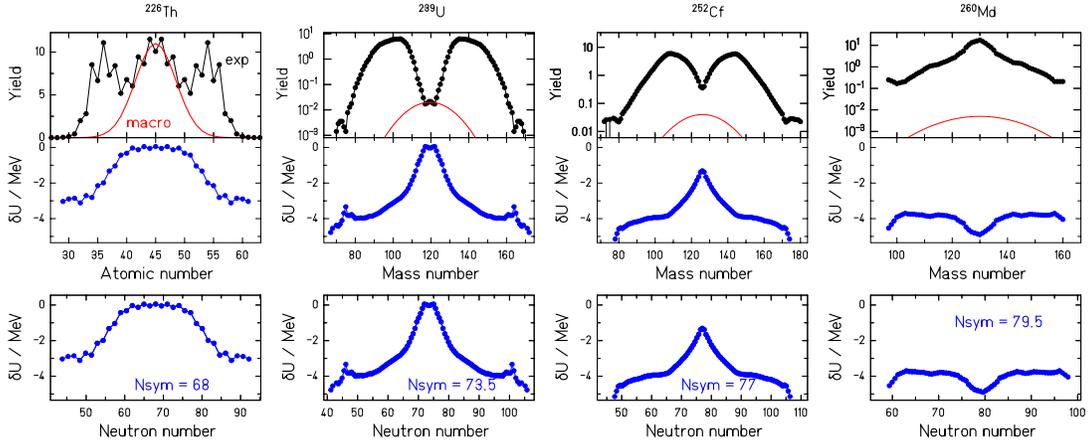

**Figure 2:** (Colour online) Extraction of the microscopic potential responsible for the nuclear-charge or mass split in fission. The amount the measured yields (data points in upper row) exceed the macroscopic prediction (red lines in upper row) is attributed to the shell-correction energy at saddle. The shell correction energies are displayed as a function of atomic number and mass number, respectively, (second row) and in a projection on neutron number (third row). See text for details. The experimental data are taken from ref. [31] ($^{226}$Th), ref. [41] ($^{239}$U), ref. [42] ($^{252}$Cf), and ref. [43] ($^{260}$Md).

At this point, we would like to cite a result of two-centre shell-model calculations, first reported by Mosel and Schmitt [18]: "By analyzing the single-particle states along the fission path ... we have established the fact that the influence of fragment shells reaches far into the PES. The preformation of the fragments is almost completed already at a point where the nuclear shape is necked in only to 40 %.". This statement suggests that the shells in the relevant configuration close to the outer saddle, determining the mass split, closely resemble the shell effects in the separate fragments.

This finding has been corroborated in later studies, but the following conclusion has never been exploited with its full weight: While the macroscopic potential, parameterised as a function of $Z^2/A$ depends on the compound nucleus, the microscopic potential is fully determined by the numbers of neutrons and protons in the nascent fragments. We name this statement the separability principle of compound-nucleus and fragment properties on the fission path. According to this principle it should be possible to trace the microscopic structures deduced in Figure 2 back to shells in the fragments, which should be the same for all systems. The success of this approach is illustrated in Figure 3 in a schematic way. Obviously, the complex behaviour of shell structure as a function of mass asymmetry of the four systems shown in Figure 2 can be reproduced already rather well as a superposition of only two shells at $N$=82 and $N$=92. The choice of these two shells is motivated by shell-model calculations, see e.g. [16,44]. The different shapes of the microscopic potentials for the four systems are explained by the different number of neutrons in the fissioning system which varies from 136 to 159. As a specific feature, in $^{260}$Md, the $N$=82 shell is approximately met in both fragments at the same time. This overlay is the reason for the strong and narrow shell effect at symmetry which leads to the appearance of the narrow symmetric fission component observed in this system.

Following the results of theoretical calculations of ref. [24], we assume for $^{226}$Th, $^{239}$U, and $^{252}$Cf, where the spherical heavy fragment is formed together with a strongly deformed light fragment, that the $N=82$ shell appears at $N=85$ in the final fragments, assuming that the heavy fragment receives 3 neutrons from the neck. This shift slightly improves the agreement with the empirical shells. For $^{260}$Md this shift is not applied, as in this case two spherical shells are formed simultaneously leading to a very compact configuration and, thus, the mass of the neck is assumed to be negligible. The $N=82$ neutron shell appears weak compared to the ground-state shell correction of $^{132}$Sn, which amounts to about -12 MeV. Two effects may be responsible for the reduction: (i) The macroscopic energy is not optimum in this fission channel, because the Coulomb energy is high due to the spherical shape of the heavy fragment. (ii) The neck perturbs the spherical symmetry of the heavy fragment. Figure 3 suggests that the deformed shell, which we assumed to be centred at $N=92$ extends to higher neutron numbers. This would indicate that the shape deviates from a Gaussian. However, we do not want to over-interpret our schematic approach, since the reduction of the complex shell structure in the neutron and proton subsystems of both fragments is certainly poorly represented by only two neutron shells.

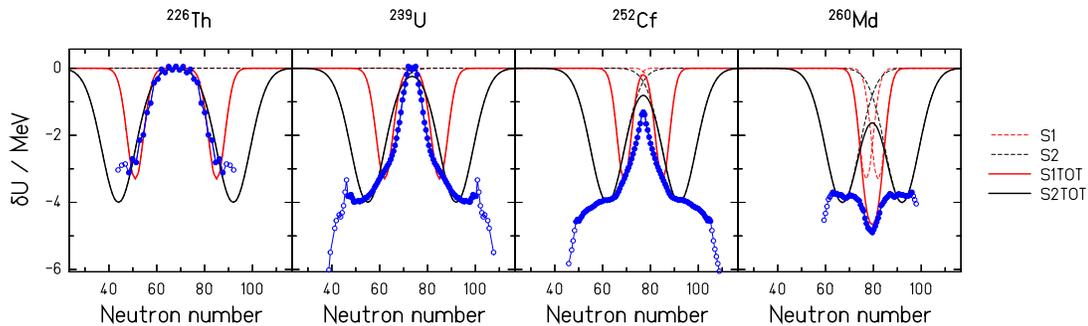

**Figure 3:** (Colour online) Comparison of the empirical shell corrections (points) and the shell-correction energies constructed from two neutron shells at $N=82$ (S1, red lines) and $N=92$ (S2, black lines). The shells are assumed to have a Gaussian shape. Position, width and depth of the shells are identical for the four nuclei (see Table 2). The sum of the corresponding shells in both fragments (S1TOT and S2TOT) is shown in addition. The deviations between the empirical values and the constructed shells at very large mass asymmetry are most probably caused by large experimental uncertainties, e.g. spurious events in the mass-yield curves (Figure 2) due to scattering. The most uncertain data are shown as open symbols.

**Table 2:** Parameters of the shells depicted in Figure 3. The same parameters were used for all fissioning systems. See text for details.

| Shell | Depth | Width ($\sigma_A$) |
|---|---|---|
| $N=82$ | -3.3 MeV | 3.5 |
| $N=92$ | -4.0 MeV | 7.0 |

*Discussion*

Although our approach is in line with many theoretical ideas and models, it disregards any influence of clustering or integrity of spherical shells [10,25,27]. Our approach is also at variance with models, which deduce fission-fragment yields from tunnelling probabilities for excitation energies well above the barrier [45,46].

The separability principle makes our approach technically rather similar to the scission-point model of Wilkins *et al.* [16]. In both cases, the yields of the fission fragments are determined by the phase space above the mass-asymmetry dependent potential, given as the sum of the macroscopic and the microscopic potential as a function of the mass split. In both cases, the microscopic potential is related to the shell effects in the separate fragments. There is, however, an important difference: While Wilkins *et al.* considered the potential energy at the scission point to govern the nuclide formation in fission, the present approach assumes that the potential in a configuration somewhere between saddle and scission is decisive for the mass split in fission. Only due to the separability principle, we assume that the shell effects of the fragments are decisive already at an earlier stage on the fission path. An important consequence is that the stiffness of the macroscopic potential at this point is considerably smaller than at scission.

Other differences of the present approach to the one of Wilkins *et al.* are that we determine the relevant properties of the potential at the fission path from experiment, that we relate the phase space to the level densities instead of using Boltzmann statistics, and that we extend our approach to energies below the barrier, where fission is governed by tunnelling.

An essential benefit of the separability principle is that the influence of shell effects on the fission process can be understood as a property of the nascent fragments. While separate calculations of shell effects or separate microscopic calculations for the different fissioning systems suffer from individual numerical uncertainties attributed to every single system and thus cause spurious fluctuations of the result as a function of mass or atomic number of the fissioning system, the separability principle suggests that the shell effects are essentially the same for all fissioning systems, since they are determined as a function of $N_{1,2}$, $Z_{1,2}$, the number of neutrons and protons in the two nascent fragments. This supports the idea of "universality" of shell corrections brought up in ref. [38]. This approach allows for a systematic view of the microscopic features of different fissioning systems. In practise, the problem of determining the microscopic component of the nuclear potential in fission reduces to establishing a systematics of microscopic features on a two-dimensional map as a function of neutron number and atomic number of the fragments.

For determining the microscopic contribution to the potential along a specific degree of freedom, e.g. mass asymmetry, one should consider that for a given mass asymmetry the shape with the maximum macroscopic binding energy differs from the optimum shape with microscopic effects included. Therefore, the energy difference between the macroscopic and the full mass-asymmetric potential at a given elongation is not just given by the microscopic energy obtained by the Strutinsky procedure. The variation of the macroscopic potential corresponding to the two different shapes must also be taken into account for evaluating the effective shell-correction energy.

For practical applications it might be advantageous to deduce the effective shell-correction energies as a function of neutron and proton number of the nascent fragments by a kind of unfolding, based on the nuclide yields of several fissioning systems. The result of this procedure would yield the effective shell-correction energies defined above. Compared to Figure 3, which demonstrates this procedure in a schematic way, a detailed analysis of the available body of experimental data might reveal the quantitative influences of more shells, e.g. $Z$=50, than just the two major ones shown in Figure 3. Semi-empirical calculations of fission-fragment nuclide yields in refs. [47,48] were based on this kind of approach.

In a strict way, our approach reduces the influence of dynamics to applying the statistical model to a configuration before reaching scission. Different degrees of freedom are assumed to freeze out at different stages. For the rest of the motion towards scission, these degrees of freedom are assumed to be frozen. By this schematic treatment the inertia and the friction tensors have no influence on the corresponding observables. Such an approach is certainly debatable [49]. It would be justified if a very slow motion in fission direction, e.g. up to the saddle, which allows for adapting the population of states in an adiabatic way, is followed by a fast motion towards scission with a time scale, which is fast compared to the characteristic time of the degree of freedom considered. This picture has proven to be valid for the angular distribution of the fission fragments in fission reactions induced by light particles [34]. It might not be true for other observables. However, if the relevant characteristics of the potential-energy surface are determined empirically as described above, some influence of the dynamics, e.g. due to structural effects in the dissipation tensor [49], may effectively be included in the deduced potential energy. This may be the reason for the values of $\hbar\omega$, which are somewhat larger than those deduced from fission excitation functions [50,51].

*Summary*

Combining a few ideas, mostly developed several decades ago leads us astonishingly far and points to the dominant features which rule the fission process. The separability of compound-nucleus and fragment properties of the system on the fission path seems to be realized to a good approximation and makes the macro-microscopic approach particularly strong in its application to nuclear fission. By deducing the shell effects from the measured fission-fragment nuclide distributions and attributing those to two major shells in the nascent fragments, we arrived at a remarkably realistic reproduction of the microscopic features of fission over the whole range covered by experiment. This approach is also suited for robust extrapolations, e.g. it has already been used to predict nuclide distributions from the fission of neutron-rich nuclei on the astrophysical r-process path [52].

In a general sense, our result also contributes to improving the understanding of the fission process. While microscopic models are indispensable for tracing back the observations as much as possible to the roots, it is equally important to reveal global tendencies and ordering principles in order to establish the hidden common features behind the complex observations.

The general importance of our result for the understanding of out-of-equilibrium processes may be summarized by the two following statements:

1. Statistical approaches might still work for out-of-equilibrium reactions, but the dynamics of the system imposes some modifications. The most important one is a memory, e.g. due to the influence of inertia on the reaction path. This influence makes it inappropriate to evaluate the relevant statistical weights of the final states at the scission point, which is the last configuration where the nascent fragments are able to interchange nucleons. Different degrees of freedom may be frozen at different positions along the fission path, depending on their time scales compared to the saddle-to-scission time.

2. The validity of the separability principle for the fission process mainly requires considering the evolution of the macroscopic potential along the fission path, while variations of the microscopic potential are less important. One may imagine that the separability principle is also found in the decay of other microscopic meta-stable systems and thus facilitates their model description. The reason is that the wave functions do not change suddenly but develop gradually towards the final configuration.